\documentclass[11pt,a4paper]{article}
\usepackage[sc,noBBpl]{mathpazo}
\usepackage[mathscr]{eucal}
\usepackage{amsmath,amsfonts,amssymb,amsthm}
\usepackage{graphicx}
\usepackage{mathtools}
\usepackage{tgpagella}
\usepackage[usenames,dvipsnames]{color}
\theoremstyle{plain}
\newtheorem{theorem}{Theorem}
\newtheorem{lemma}[theorem]{Lemma}

\newtheoremstyle{note}{\topsep}{\topsep}{\slshape}{}{\scshape}{}{ }{}
\theoremstyle{note}

\newtheoremstyle{warning}{\topsep}{\topsep}{\color{Blue}\small\slshape}{}{\color{Red}\scshape}{.}{ }{}
\theoremstyle{warning}

\numberwithin{equation}{section}
\numberwithin{theorem}{section}
\renewcommand{\theparagraph}{\S\,\arabic{paragraph}}
\usepackage[sc,small,pagestyles]{titlesec}

\titleformat{\paragraph}[runin]
  {\normalfont\bfseries}
  {\theparagraph.}{.5em}{}[.]
\titlespacing{\paragraph}
  {0pt}{1.5ex plus .1ex minus .2ex}{1em}
\newcommand\be{{\mathbf e}}

\newcommand\br{{\mathbf r}}

\newcommand\bz{{\mathbf z}}
\newcommand\bzero{{\mathbf 0}}

\newcommand\bA{{\mathbf A}}

\newcommand\bE{{\mathbf E}}

\newcommand\bJ{{\mathbf J}}

\newcommand\bP{{\mathbf P}}

\newcommand\bR{{\mathbf R}}

\newcommand\bU{{\mathbf U}}
\newcommand\bV{{\mathbf V}}

\newcommand\bZ{{\mathbf Z}}
\newcommand\cD{{\mathcal D}}

\newcommand\cG{{\mathcal G}}

\newcommand\cT{{\mathcal T}}

\newcommand\scN{{\mathscr N}}

\newcommand\field{\mathbb}

\newcommand\R{\field{R}}
\newcommand\bbS{\mathbb{S}}

\newcommand\C{\field{C}}

\newcommand\bGamma{\boldsymbol{\Gamma}}

\newcommand\rmi{\mathrm{i}\mspace{1mu}}

\newcommand\Dt{\frac{\mathrm{d}\phantom{t} }{\mathrm{d}\mspace{1mu}
t}}

\newcommand\Dz{\frac{\mathrm{d}\phantom{z} }{ \mathrm{d}z}}

\newcommand\pder[2]{\dfrac{\partial #1 }{\partial #2}}

\newcommand\DDz{\frac{\mathrm{d}^2\phantom{z} }{ \mathrm{d}z^2}} 
\newcommand\DDt{\frac{\mathrm{d}^2\phantom{t} }{\mathrm{d}\mspace{1mu}
t^2}}
\newcommand\norm[1]{\lVert #1 \rVert}

\newcommand\mtext[1]{\quad\text{#1}\quad}

\newcommand\defset[2]{\left\{{#1}\;\vert \;\; {#2} \,\right\}}

\title{Non-integrability of the dumbbell and point mass problem}
\author{
  Andrzej J.~Maciejewski$^1$, Maria Przybylska$^2$,  \\ Leon Simpson$^1$, 
   and  Wojciech Szumi\'nski$^2$ \\[1em]
  {}$^1$Kepler Institute of Astronomy,\\ University of Zielona G\'ora, 
  Licealna 9,  \\
  PL-65-407,  Zielona G\'ora, Poland,  \\[1em]
  {}$^{2}$Institute of Physics, \\ University of Zielona G\'ora, 
  Licealna 9,  \\
  PL-65-407,  Zielona G\'ora, Poland }
\date{\today}

\begin{document}
\mathtoolsset{%
  mathic,centercolon%
}
\maketitle

\begin{abstract}
  This paper discusses a constrained gravitational three-body problem
  with two of the point masses separated by a massless inflexible rod
  to form a dumbbell. The non-integrability of this system is proven
  using differential Galois theory.
\end{abstract}

\section{Equations of motion, symmetries and reduction}
\begin{figure}[ht]
  \begin{center}
    \includegraphics[scale=0.7]{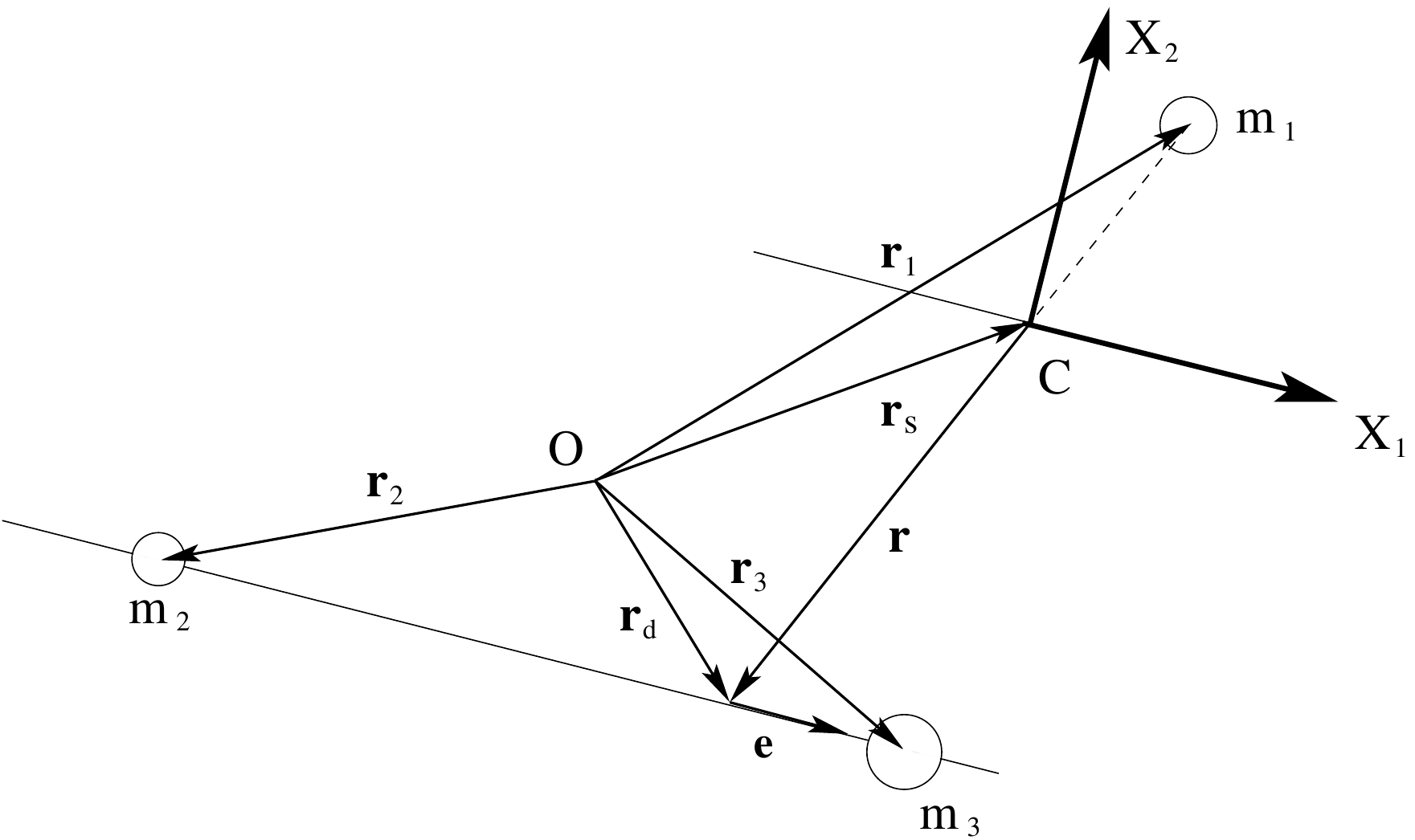}
  \end{center}
  \caption{\label{fig:geo} A dumbbell and a point mass in the plane. Point $O$ denotes the origin of an inertial frame. Point $C$ is the system centre of masses. Axis $OX_1$ of the moving frame is always parallel to the dumbbell.}
\end{figure}

Considered is the gravitational three-body problem with a single constraint. Three point masses, $m_1$, $m_2$ and $m_3$ move in a plane under mutual gravitational interaction. Masses $m_2$ and $m_3$ are
connected by a massless inflexible rod of length $l>0$ to form a dumbbell. A pictorial description of the problem is given in Figure~\ref{fig:geo}.

The Lagrangian of the system has the form 
\begin{equation}
  L = \frac{1}{2} m_1\Vert {\dot{\br}}_{1}\Vert^2 + \frac{1}{2} m_2
  \Vert{\dot{\br}}_{2}\Vert^2 + \frac{1}{2} m_3\Vert {\dot{\br}}_3\Vert^2 -
  U(\br_1,\br_2,\br_3),
\end{equation}
where $\br_1$, $\br_2$ and $\br_3$ are the position vectors of the
respective masses, and
\begin{equation}
  U(\br_1,\br_2,\br_3) = - \frac{G m_1
    m_2}{\Vert\br_1-\br_2\Vert } - \frac{G m_1
    m_3}{\Vert \br_1-\br_3\Vert } - \frac{G m_2
    m_3}{\Vert \br_2-\br_3\Vert },
\end{equation}
is the potential. $\br_{\mathrm{d}}$ is the radius vector of the centre of mass of the dumbbell, and
\begin{equation}
  \label{eq:1}
  \be:= \frac{\br_3-\br_2}{\Vert \br_3-\br_2 \Vert},
\end{equation}
is a unit vector along the dumbbell. It follows that
\begin{align}
  \br_2 &= \br_\mathrm{d} + \mu_3 l \be, \qquad
  {\dot\br}_2 = {\dot\br}_\mathrm{d} + \mu_3 l \dot \be, \\
  \br_3 &= \br_\mathrm{d} - \mu_2 l \be, \qquad {\dot\br}_3 =
  {\dot\br}_\mathrm{d} - \mu_2 l \dot\be,
\end{align}
where
\begin{equation}
  \label{eq:2}
  \mu_2:=  \frac{m_2}{m_{\mathrm{d}}}, \qquad
  \mu_3:=  \frac{m_3}{m_{\mathrm{d}}}, \qquad 
  m_{\mathrm{d}}:= m_2 + m_3.
\end{equation}
The system has five degrees of freedom. The configuration space of the point $m_1$ is $\R^2$, and the configuration space of the dumbbell is $\R^2\times \bbS^{1}$. A configuration of the system is fully specified by $\br_1$, $\br_{\mathrm{d}}$, and $\be$. In these coordinates, the Lagrangian is as follows.
\begin{equation}
  \label{eq:3}
  L = \frac{1}{2} m_1 \norm{{\dot\br}_1}^2 +
  \frac{1}{2} m_{\mathrm{d}}\norm{{\dot\br}_\mathrm{d}}^2 + 
  \frac{1}{2}I \norm{{\dot\be}}^{2} -\widetilde U(\br_1, \br_{\mathrm{d}}, \be), 
\end{equation}
where
\begin{equation}
  \label{eq:4}
  I:= l^2 \frac{m_2m_3}{m_{\mathrm{d}}}, \qquad 
  \widetilde U(\br_1, \br_{\mathrm{d}}, \be)=
  U(\br_1, \br_\mathrm{d} + \mu_3 l \be,  \br_\mathrm{d} - \mu_2 l \be).
\end{equation} 
The system possesses natural symmetries that are exploited to reduce the dimension of the system. Amongst these is translational symmetry. This is manifest by defining $\br:= \br_{\mathrm{d}}-\br_1$ as the vector between the mass $m_1$ and the dumbbell. A configuration of the system can be thus described
by $\br$, $\be$, and $\br_{\mathrm{s}}$ defined by
\begin{equation}
  \label{eq:5}
  m_\mathrm{s} \br_{\mathrm{s}}=m_1 \br_1 + m_{\mathrm{d}}\br_{\mathrm{d}},
  \qquad  m_\mathrm{s}:= m_1+m_2+m_3.
\end{equation}
As
\begin{equation}
  \label{eq:6}
  \br_1= \br_{\mathrm{s}} -\frac{m_{\mathrm{d}}}{m_{\mathrm{s}}}\br, \qquad
  \br_{\mathrm{d}}=\br_{\mathrm{s}} +\frac{m_1}{m_{\mathrm{s}}}\br, 
\end{equation}
the Lagrangian \eqref{eq:3} is written in the following form.
\begin{equation}
  \label{eq:7}
  L= \frac{1}{2} m_{\mathrm{r}} \Vert{\dot\br}\Vert^2 + 
  \frac{1}{2}I\Vert {\dot\be}\Vert^2 +
  \frac{1}{2} m_{\mathrm{s}}\Vert{\dot\br}_\mathrm{s}\Vert^2 - W(\br, \be), 
\end{equation}
where
\begin{equation}
  \label{eq:8}
  m_\mathrm{r} := \frac{m_1 m_{\mathrm{d}}}{m_{\mathrm{s}}},
\end{equation}
is the reduced mass, and
\begin{equation}
  \label{eq:9}
  W(\br,\be)=  
  - \frac{G m_1 m_2}{\Vert\br +\mu_3 l \be \Vert } - \frac{G m_1
    m_3}{\Vert \br -\mu_2 l \be\Vert }.
\end{equation}
The components of $\br_{\mathrm{s}}$ are cyclic coordinates, and the motion of the centre of mass separates completely. Therefore, the term $m_{\mathrm{s}}\Vert{\dot\br}_\mathrm{s}\Vert^2/2$ is removed from the Lagrangian~\eqref{eq:7}.

It is convenient to introduce dimensionless variables, taking $l$ as the unit of length, $m_{\mathrm{r}}$ as the unit of mass, and
\begin{equation}
  \label{eq:16}
  T:=\sqrt{\frac{l^3}{m_{\mathrm{s}}}},
\end{equation}
as the unit of time. Setting $l=m_{\mathrm{r}}=T=1$, the Lagrangian~\eqref{eq:7} reduces to
\begin{equation}
  \label{eq:17}
  L= \frac{1}{2} \Vert {\dot\br}\Vert^2 + 
  \frac{1}{2}I \Vert{\dot\be}\Vert^2 
  - W(\br, \be),
\end{equation}
where
\begin{equation}
  \label{eq:18}
  W(\br,\be)=  
  - \frac{1}{\Vert\left( 1+\mu \right)\br +\mu \be \Vert } - 
  \frac{\mu}{\Vert \left( 1+\mu \right)\br -\be\Vert },
  \qquad \mu:=\frac{m_3}{m_2}.
\end{equation}
The reduced system still has a symmetry: it is invariant with respect to the natural action of the group $\mathrm{SO}(2,\R)$. In fact, the dynamics are oblivious to the orientation of the inertial frame. This
symmetry can be used to reduce the dimension of the configuration space by one. This is achieved by describing the dynamics in a rotating frame in which the dumbbell is at rest. The transformation from the inertial frame to the rotating frame is given by 
\begin{equation}
  \label{eq:11}
  \br = \bA \bR,  \qquad \bA \in \mathrm{SO}(2,\R). 
\end{equation}
An additional assumption, that $\be$ is parallel to $x$-axis of the rotating frame, allows for a coordinate representation of the transformation as follows.
\begin{equation}
  \label{eq:10}
  \bA= \begin{bmatrix*}[r] 
    \cos\varphi & -\sin\varphi \\
    \sin\varphi &\cos\varphi
  \end{bmatrix*}, \mtext{and} \be=[\cos\varphi,\sin\varphi]^T. 
\end{equation}
Then 
\begin{equation}
\label{eq:e}
\be=\bA\bE, \qquad \bE = [1,0]^T,
\end{equation}
and
\begin{equation}
  \label{eq:12}
  \dot \be = \dot \varphi [-\sin\varphi, \cos\varphi]^T , \qquad 
  \dot \bA= \dot\varphi\bA\bJ, \qquad 
  \bJ = \begin{bmatrix*}[r]
    0 & -1 \\
    1 & 0
  \end{bmatrix*}.
\end{equation}
The Lagrangian expressed in the coordinates $\bR=[X_1, X_2]^T$, and $\varphi$ has the form
\begin{equation}
  \label{eq:13}
  L = \frac{1}{2}{\Vert\dot\bR\Vert}^2 +\dot\varphi {\dot\bR}^T\bJ\bR +
  \frac{1}{2}{\dot\varphi}^2 \Vert\bR\Vert^{2} +\frac{1}{2}I {\dot\varphi}^2 - V(\bR), 
\end{equation}
where
\begin{equation}
  \label{eq:19}
  V(\bR):=  
  - \frac{1}{\Vert\left( 1+\mu \right)\bR +\mu \bE \Vert } - 
  \frac{\mu}{\Vert \left( 1+\mu \right)\bR -\bE\Vert },
\end{equation}
The generalised momenta $\bP:=\left[ P_1,P_2 \right]^T$ and $P_{\varphi}$, are given by
\begin{equation}
  \label{eq:P}
  \bP:= \pder{L}{\dot\bR}=\dot\bR +\dot\varphi\bJ\bR, \qquad
  P_{\varphi}:= \pder{L}{\dot\varphi}=\dot\varphi \left(\Vert \bR\Vert^2+ I\right)  + {\dot\bR}^T\bJ\bR. 
\end{equation}
Coordinate $\varphi$ is cyclic, so $P_{\varphi}$ is a first integral of the system. Thus, the Hamiltonian of the system is written as
\begin{equation}
  \label{eq:H}
  H=\frac{1}{2}\Vert \bP \Vert^{2 }+\frac{1}{2}\alpha \left(\gamma - \bP^T\bJ\bR \right)^2 +V(\bR), 
\end{equation}
where $\alpha=1/I$, and $\gamma$ is a fixed value of the first integral $P_{\varphi}$ corresponding to the cyclic variable $\varphi$. Hence, the equations of motion have the form
\begin{equation}
  \label{eq:em}
  \begin{split}
    \Dt \bR = &\bP -\alpha \left( \gamma -\bP^T\bJ\bR \right)\bJ\bR, \\
    \Dt \bP = &- \alpha \left( \gamma -\bP^T\bJ\bR \right)\bJ\bP-
    \left( 1+\mu \right)
    \frac{\left( 1+\mu \right)\bR +\mu \bE}{\Vert\left( 1+\mu \right)\bR +\mu \bE \Vert^3} - \\
    & \mu \left( 1+\mu \right)\frac{\left( 1+\mu \right)\bR
      -\bE}{\Vert \left( 1+\mu \right)\bR -\bE\Vert^3 }.
  \end{split}
\end{equation}
\section{Problem}
\label{sec:pro}
Numerical experiments suggest that the considered system is not integrable. The complex behaviour of the system is apparent from the Poincar\'e cross-section. For fixed values of the parameters and the
energy the equations~\eqref{eq:em} are integrated numerically by the Burlish-Stoer method  \cite{Press:92::}. The cross-section plane is specified as $X_1=0$, and the cross direction is chosen $\dot X_1>0$. The coordinates of the cross-section are $(X_2,P_2)$. The cross-sections presented in the subsequent figures are obtained for parameters $\mu=1$ and $\alpha=\frac{4}{61}$.

\begin{figure}[ht]
  \begin{center}
   \includegraphics[scale=0.25]{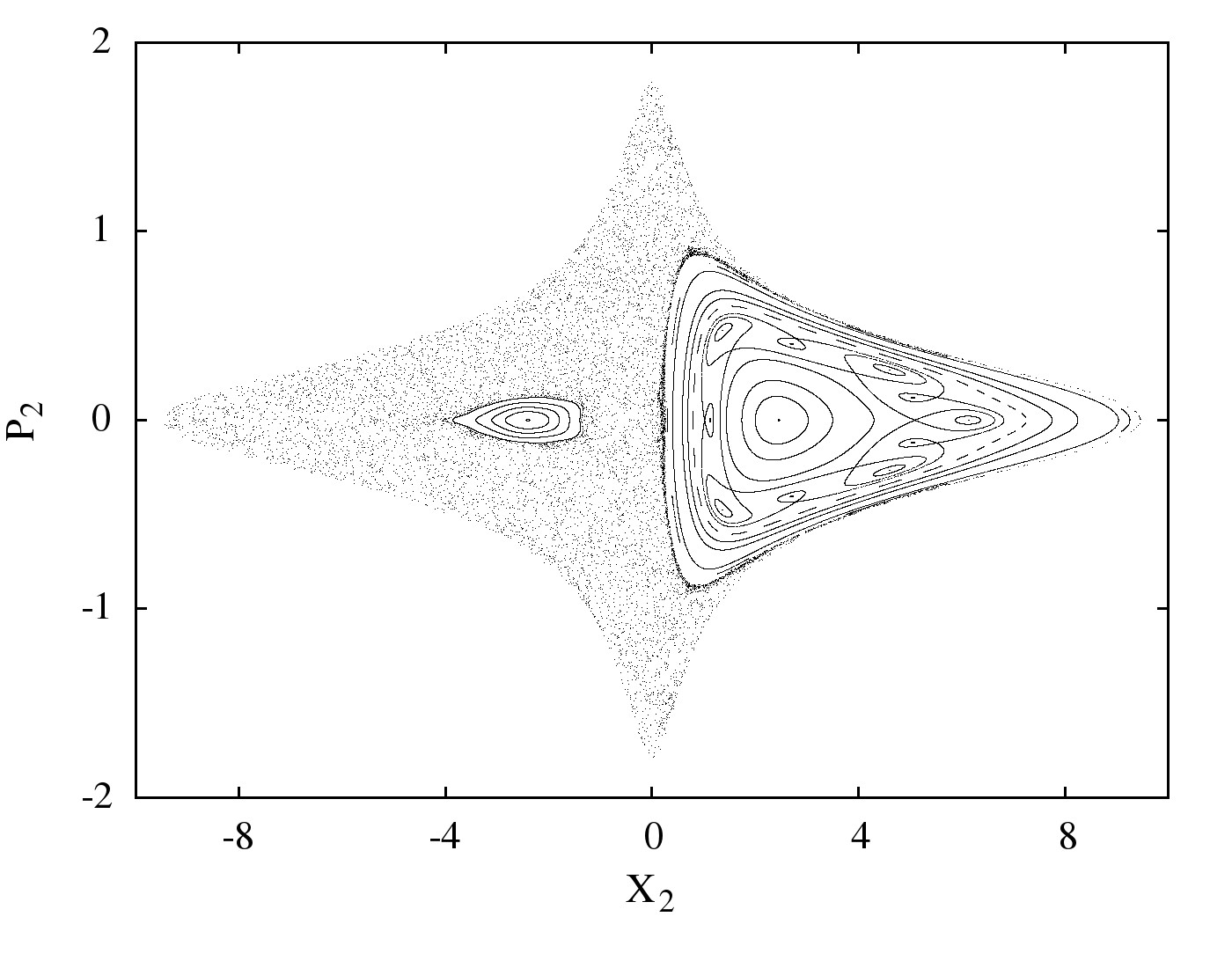}
  \end{center}
  \caption{\label{fig:a1} Cross-section for energy $e=-0.1$ and $\gamma=-1$}
\end{figure}

\begin{figure}[ht]
  \begin{center}
    \includegraphics[scale=0.25]{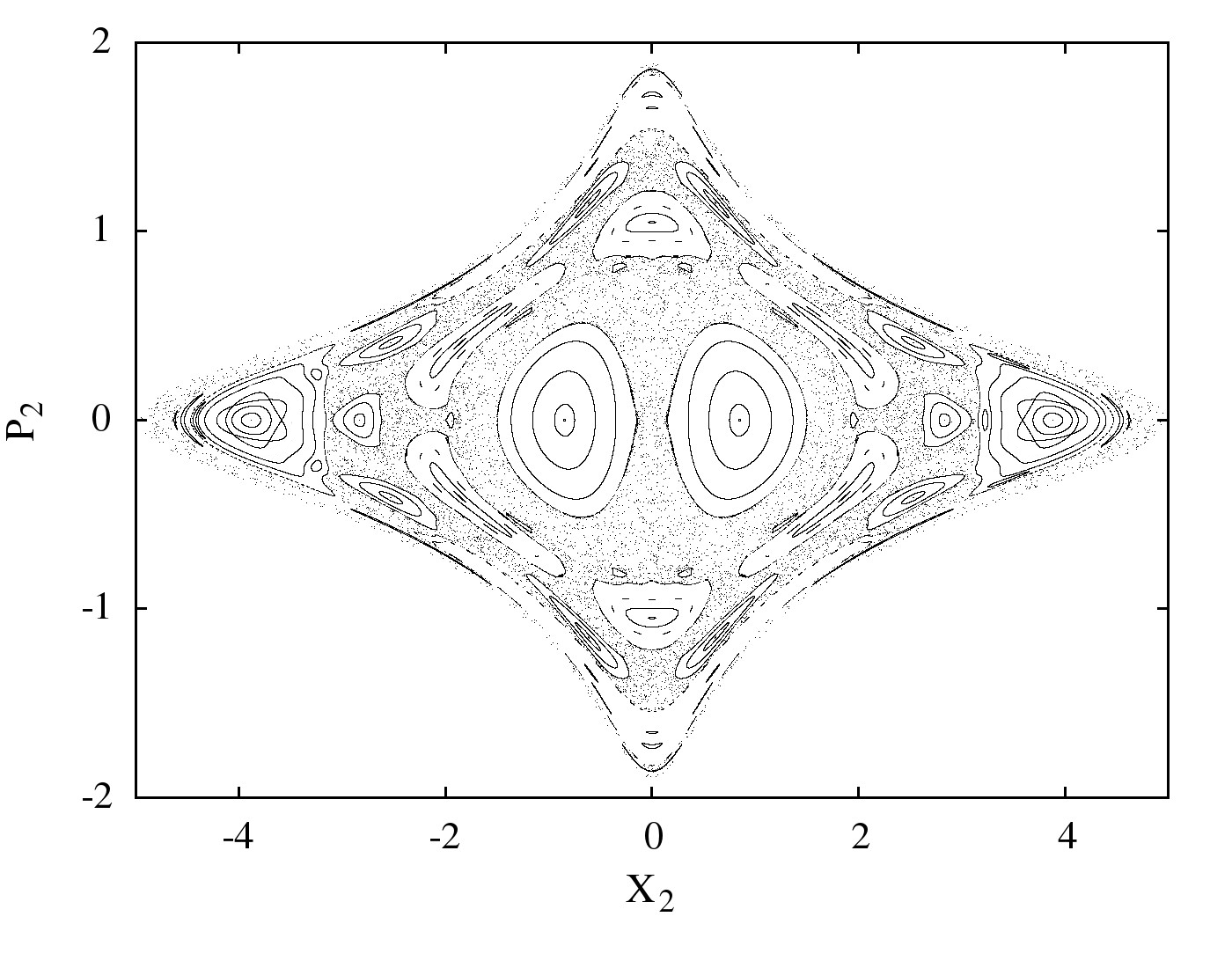}
  \end{center}
  \caption{\label{fig:a2} Cross-section for $e=-0.2$ and $\gamma=0$}
\end{figure}

\begin{figure}[ht]
  \begin{center}
   \includegraphics[scale=0.25]{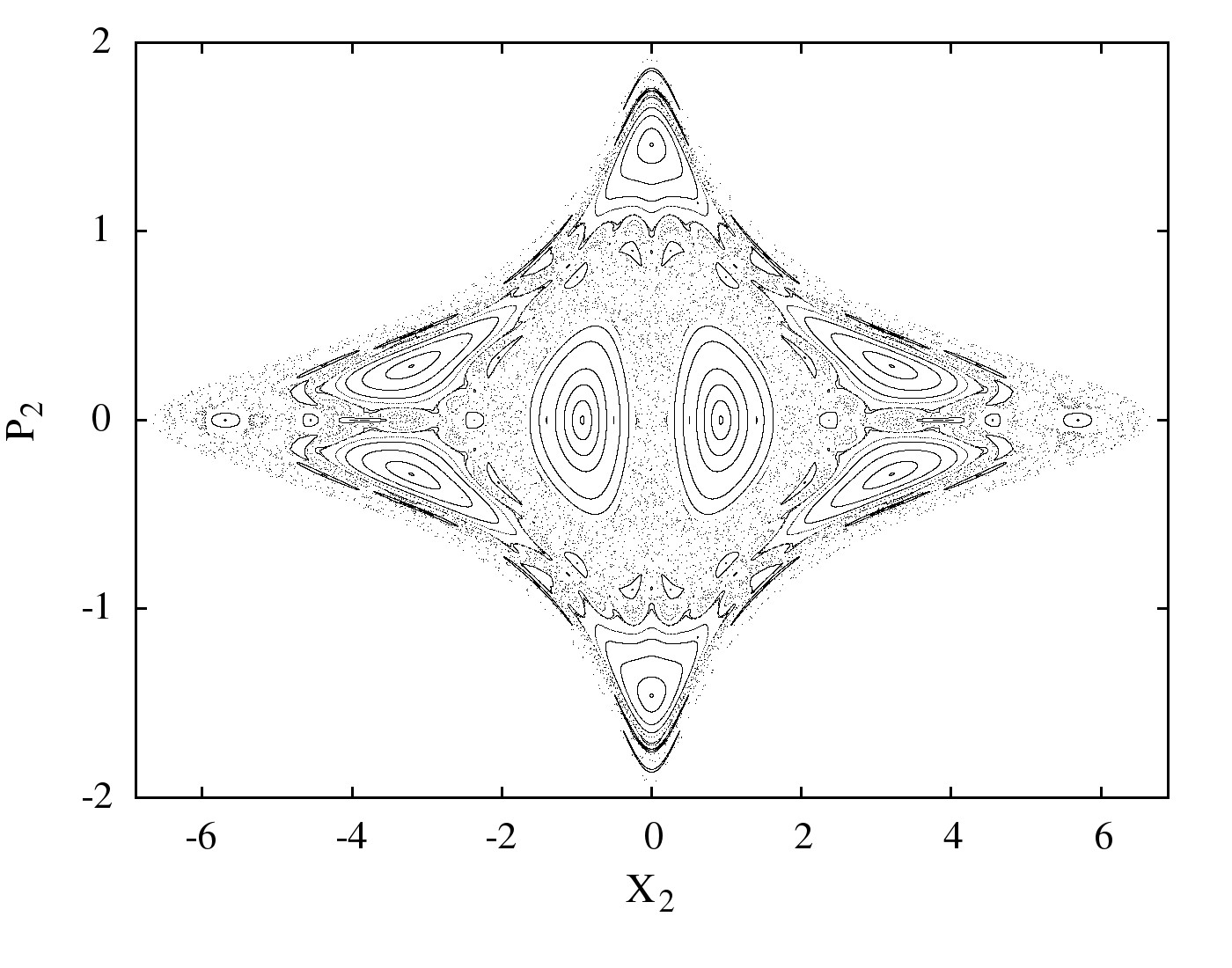}
  \end{center}
  \caption{\label{fig:a3} Cross-section for $e=-0.15$ for system with $\gamma=0$}
\end{figure}

Figure~\ref{fig:a1} is a cross-section corresponding to energy $e= -0.1$ and $\gamma=-1$. Chaotic behaviour is evident.  
Two cross-sections for $\gamma=0$ are shown in Figure~\ref{fig:a2} and Figure~\ref{fig:a3}. Both of them attest to the non-integrability of the system.
The problem is proving the non-integrability of the system analytically. It is attractive to try using the differential Galois approach to the problem of integrability (see e.g.~\cite{Morales:99::c}). However, as is explained in \cite{Combot:13::}, application of this theory directly to the system~\eqref{eq:em} is invalid as the Hamiltonian function \eqref{eq:H} is not single-valued.

Proposed in this paper is a solution addressing this deficiency, based on an extension of the system into  a larger phase space. Coordinates of these additional dimensions are denoted by $R_1$ and $R_2$. This approach further requires a definition of the time derivative consistent with system~\eqref{eq:em}. Following this, coordinates $R_1$ and $R_2$ are considered as lengths of vectors 
\begin{equation}
  \label{eq:27}
  \bR_1:= \left( 1+\mu \right)\bR +\mu \bE, \qquad \bR_2:= \left( 1+\mu \right)\bR -\bE. 
\end{equation}
That is, equations $G_1=G_2=0$, where polynomials $G_1, G_2 \in\C[X_1, X_2, R_1, R_2]$ given by 
\begin{equation}
  \label{eq:15}
  \left.
    \begin{aligned}
      G_1:=&R_1^2-\left[ \left( 1+\mu)X +\mu \right) \right]^2-\left( 1+\mu \right)^2Y^2=R_1^2-\bR_1^T\bR_1,\\
      G_2:=&R_2^2-\left[ \left( 1+\mu)X -1\right) \right]^2-\left(
        1+\mu \right)^2Y^2=R_2^2-\bR_2^T\bR_2.
    \end{aligned}
    \quad\right\}
\end{equation}
defines $R_1$ and $R_2$ as algebraic functions of $(X_1,X_2)$.

The extended phase space is $\C^6$ with coordinates $\bZ:=\left[X_1,X_2,P_1,P_2,R_1,R_2 \right]$. The Hamiltonian~\eqref{eq:H} expressed in these coordinates reads  
\begin{equation}
\label{eq:K}
K = K(\bZ):= \frac{1}{2}\Vert \bP \Vert^{2 }+\frac{1}{2}\alpha \left(\gamma - \bP^T\bJ\bR \right)^2 -\frac{1}{R_1}-\frac{\mu}{R_2}.
\end{equation} 
As 
\begin{equation}
\label{eq:dotRi}
\Dt R_i= \frac{\bR_i^T\dot\bR_i}{R_i}=(1+\mu) \frac{\bR_i^T\dot\bR}{R_i}, \qquad i=1,2, 
\end{equation}
system~\eqref{eq:em} can be extended to the following one.
\begin{equation}
\label{eq:dotZ}
\Dt \bZ = \bU(\bZ), \qquad \bU(\bZ):=\bJ(\bZ) K'(\bZ), \qquad \bJ(\bZ):=
\begin{bmatrix*}[r]
\bzero_{2} & \bE_2 & \bzero_{2}\\
-\bE_2 & \bzero_2  & -\bA \\
\bzero_2 & \bA^T & \bzero_2
\end{bmatrix*}
\end{equation}
where 
\begin{equation}
\label{eq:29}
\bA = (1+\mu)\left[\frac{\bR_1}{R_1}, \frac{\bR_2}{R_2}  \right], \qquad K'(\bZ):=\left( \nabla_{\bZ}K(\bZ) \right)^T.
\end{equation}
It can be shown that matrix $\bJ(\bZ)$ defines a Poisson structure on $\C^6$. The corresponding Poisson bracket is given by the following formula
\begin{equation}
\label{eq:bra}
\left\{ F, G \right\}(\bZ)= F'(\bZ)^T \bJ(\bZ) G'(\bZ), 
\end{equation}
where $F$ and $G$ are two smooth functions. The Poisson structure given by $\bJ(\bZ)$ is degenerated and has rank 4. It is demonstrable that polynomials $G_1$ and $G_2$ are Casimir functions of this structure, whose common levels are symplectic manifolds. In particular,
\begin{equation}
  \label{eq:M4}
  M^4_{\mu} :=\defset{\bZ\in\C^6}{G_1(\bZ)=G_2(\bZ)=0}, \qquad \mu\in (0,1/2].
\end{equation}
is a symplectic manifold. System~\eqref{eq:dotZ}, when restricted to $ M^4_{\mu}$, is equivalent to~\eqref{eq:em}. 

In the setting thus described the main result of this paper can be formulated in the following theorem. 
\begin{theorem}
  For $\gamma=0$ and $\mu\in(0,1/2]$, system~\eqref{eq:dotZ}, when restricted to the symplectic leaf $M_{\mu}^{4}$, is not integrable in the Liouville sense with meromorphic first integrals.
\end{theorem}

\section{Particular solution and variational equation}
If $\gamma=0$, then equations~\eqref{eq:dotZ} have an invariant algebraic manifold,
\begin{equation}
  \label{eq:N}
  \scN:=\defset{\bZ\in\C^6}{X_2=P_2=0, \quad G_1(\bZ)=G_2(\bZ)=0}.
\end{equation}
On this manifold it can be taken that
\begin{equation}
\label{eq:r12}
R_1 = (1+\mu)X_1 +\mu , \qquad R_2 = (1+\mu)X_1 -1,
\end{equation}
so the system~\eqref{eq:dotZ}, restricted to $\scN$, reduces to
\begin{equation}
  \dot X=P,\qquad \dot P=-(\mu+1)\left(\dfrac{\mu}{\left[\left(1+\mu  \right)X
        -1\right]^2}+\dfrac{1}{\left[\left( 1+\mu \right)X+\mu\right]^{2}}\right),
  \label{eq:partsol}
\end{equation} 
where $(X,P):=(X_1,P_1)\in\C^2$. This is a Hamiltonian system with one degree of freedom. Thus the manifold $\scN$ is foliated by phase curves $\Gamma_{e}$ on energy levels $K_{|\scN}=e$, that is 
\begin{equation}
  \label{eq:14}
  \Gamma_e \subset M_e:=\defset{(X,P)\in \C^2}{e=\frac{1}{2}P^2- \dfrac{\mu}{\left(1+\mu  \right)X
      -1}-\dfrac{1}{\left( 1+\mu \right)X+\mu}}.
\end{equation}
For a generic value of $e$ the level $M_e$ contains three phase curve. If $e$ is real, two or three of these levels have a non-empty intersection with the real part of the phase space. In further consideration it is assumed that $e\geq 0$, and the chosen phase curve $\Gamma_{e}$ contains the half-line $(1/(1+\mu),\infty)\times \{0\}\subset\C^2$. In the phase space $\C^6$ this curve is given by 
\begin{equation}
\label{eq:pc}
\bGamma_e:=\defset{\bZ\in\scN}{ (X_1,P_1)\in \Gamma_e}.
\end{equation}

Application of the Morales-Ramis theory necessitates the linearisation of equations~\eqref{eq:dotZ} along $\Gamma_{e}$. It has the form 
\begin{equation}
\label{eq:var} 
\Dt \bz = \bV \bz , \qquad \bV=\pder{\bU}{\bZ}(\bZ), \quad \bZ\in\bGamma_e. 
\end{equation}
This system has three first integrals 
\begin{equation}
\label{eq:varz} 
k(\bz):= \nabla K(\bZ)\cdot \bz, \qquad g_1(\bz):=\nabla G_1(\bZ)\cdot\bz, \qquad g_2(\bz):=\nabla G_2(\bZ)\cdot\bz,\quad \bZ\in\bGamma_e.
\end{equation}
On the level $g_1(\bz)=g_2(\bz)=0$, the following equalities are satisfied
\begin{equation}
\label{eq:zl}
R_1 r_1 = (1+\mu)\left[(1+\mu)X_1 +\mu\right]x_1, \qquad R_2 r_2 = (1+\mu)\left[(1+\mu)X_1 -1\right]x_1, 
\end{equation} 
where it is assumed that $\bz=\left[ x_1, x_2, p_1, p_2, r_1, r_2\right]^T$. On $\bGamma_e$, coordinates $(R_1,R_2)$ can be expressed by $X_1$,
\begin{equation}
\label{eq:R12X1}
R_1= (1+\mu)X_1 +\mu, \qquad R_2=(1+\mu)X_1 -1. 
\end{equation}
Hence, on the level $g_1(\bz)=g_2(\bz)=0$, variables $(r_1, r_2)$ can be eliminated from the variational equations~\eqref{eq:var}. The reduced system of variational equations has the form
\begin{equation}
  \Dt \begin{bmatrix}
    u_1\\
    v_1\\
    u_2\\
    v_2
  \end{bmatrix}=
  \begin{bmatrix*}[r]
    0&1&0&0\\
    A & 0&0& 0 \\
    0& 0 &-\alpha X P &1 +\alpha X^2\\
    0&0 &B&\alpha X P
  \end{bmatrix*}
  \begin{bmatrix}
    u_1\\
    v_1\\
    u_2\\
    v_2
  \end{bmatrix},
\end{equation}
where
\begin{equation}
  \begin{split}
    &A=2(\mu+1)^2\left(\dfrac{\mu}{\left[ \left(1+\mu\right)
          X-1\right]^3}+\dfrac{1}{\left[\left(1+\mu\right) X+\mu\right]^3}\right),\\
    &B=-\alpha P^2-\frac{1}{2}A,
  \end{split}
\end{equation} 
and $(X,P)\in\Gamma_{e}$. This system splits into two subsystems. Variables $(u_1,v_1)$ describe variations in the invariant plane $\scN$. In fact, $(u_1,v_1)$ is a vector tangent to $\scN$ at
point $(X,P)$, hence the subsystem in variables $(u_1,v_1)$ is called \emph{tangential}. The second subsystem, corresponding to variables $(u_2,v_2)$, yield the \emph{normal variational equations}. The
latter subsystem can be expressed by a single second-order equation,
\begin{equation}
  \ddot u+a \dot u+bu=0,
  \label{eq:variaty}
\end{equation}
where $u=u_2$, and
\[
a:=-\dfrac{2 \alpha X P}{1 + \alpha X^2 },\qquad b:=\dfrac{2
  \alpha P^2}{1 + \alpha X^2 }+ \frac{1}{2} \left(1+\alpha X^2\right)
A.
\]

Changing the independent variable by
\begin{equation}
  t\longrightarrow z=(1+\mu)X
  \label{eq:ratio}
\end{equation}
transforms equation~\eqref{eq:variaty} into an equation with rational coefficients. 
This follows from
\[
\begin{split}
  &\Dt=\dot z\Dz,\qquad \DDt=(\dot z)^2\DDz+\ddot z \Dz,\\
  &\dot z=(1+\mu) \dot x=(1+\mu) p_1,\qquad (\dot
  z)^2=(1+\mu)^2
  p_1^2=2(1+\mu)^2\left(e+\dfrac{\mu}{z-1}+\dfrac{1}{z+\mu}
  \right),\\
  &\ddot z=(1+\mu)\ddot
  p_1=-(\mu+1)^2\left(\dfrac{\mu}{(z-1)^2}+\dfrac{1}{(z+\mu)^2}\right),
\end{split}
\]
where the energy integral 
\[
e=\frac{1}{2}{\dot X}^2- \dfrac{\mu}{\left(1+\mu  \right)X
      -1}-\dfrac{1}{\left( 1+\mu \right)X+\mu}
\]
has been used.

Resultantly, equation~\eqref{eq:variaty} transforms into
\begin{equation}
  X''+p(z)X'+q(z)X=0, ''\equiv\Dz,\quad X\equiv X_1,
  \label{eq:zvar}
\end{equation} 
with
\[
\begin{split}
  p&=-\left[(-1 + z)^2 (1 + z^2 (5 + 4 e z) \alpha) + (2 + z (-4 + 3 z
    + (4 +
    z (-8 + 8 e (-1 + z)^2\right.\\
  &\left. + 5 z^2)) \alpha)) \mu + (1 + 3 z^2 + 2 z (2 e (-1 + z)^2 +
    z (-4 + 5 z)) \alpha) \mu^2 + (1 +
    z (4 + z - 4 \alpha\right.\\
  &\left. + 5 z \alpha)) \mu^3 + 2 (1 + z) \mu^4 +
    \mu^5\right]/\left[2 (z-1) (z + \mu) ((z-1) (1 + e z) + (e (z-1) +
    z) \mu\right.\\
  &\left. + \mu^2) (z^2 \alpha + (1 + \mu)^2)\right],\\
  q&=\left[4 (z-1)^2 \alpha (1 + \mu) (z + \mu)^2 ((z-1) (1 +
    e z) + (e (z-1) + z) \mu + \mu^2) + ((z-1)^3\right.\\
  &\left. + (1 + 3 (z-1) z) \mu + (3 z-1) \mu^2 + \mu^3) (z^2 \alpha +
    (1 + \mu)^2)^2\right]/\left[2
    (z-1)^2 (1 + \mu)\right.\\
  &\left.\cdot (z + \mu)^2 ((z-1) (1 + e z) + (e (z-1) + z) \mu +
    \mu^2) (z^2 \alpha + (1 + \mu)^2)\right]
\end{split}.
\]

Making the next change of variables,
\begin{equation}
  \label{eq:tran}
  X = w \exp\left[ -\frac{1}{2} \int_{z_0}^z p(s)\, ds \right]
\end{equation}
\eqref{eq:zvar} is transformed to the canonical form,
\begin{equation}
  \label{eq:normal}
  w'' = r(z) w, \qquad r(z) = -q(z) + \frac{1}{2}p'(z)  + \frac{1}{4}p(z)^2,
\end{equation}
with coefficients given as
\begin{equation}
  \begin{split}
    & r=\dfrac{P}{M},\\
    &M=16 (z-1)^2 \beta^2 (z^2 + \beta^2)^2 (z + \mu)^2 [(z-1) (1 + e
    z) + (e (-1
    + z) + z) \mu + \mu^2]^2,\\
    &P=\sum_{i=0}^{11}p_iz^{11-i}.
  \end{split}
  \label{eq:rr}
\end{equation} 
Here $\beta^2=(1+\mu)^2/\alpha$ and coefficients $p_i$ of $P$ are given in the Appendix in \eqref{eq:coeffy}. This equation has seven singularities,
\[
\begin{split}
  &z_1=0,\qquad z_2=-\mu,\qquad  z_{3,4}=\pm \rmi\beta,\\
  & z_{5,6}= \dfrac{e-1-(1 + e) \mu\pm\sqrt{(1 + \mu) ((1 + e)^2 +
      (e-1)^2 \mu)}}{2e},
\end{split}
\]
and $z_7=\infty$, providing energy is chosen such that
\begin{equation}
  \begin{split}
    &e\not\in\left\{ \dfrac{(1 \pm \rmi \beta) (\pm\rmi + \beta \mp
        \rmi \mu) (\beta \pm \rmi \mu) (1 + \mu)}{(1 + \beta^2)
        (\beta^2 +
        \mu^2)},\dfrac{(\sqrt{\mu}\pm\rmi)^2}{\mu+1},\right.\\
    &\left.\dfrac{-\mu + \mu^3 \pm\rmi (1 + \mu^3 + \beta^2 (1 +
        \mu))|\beta|}{(1 + \beta^2) (\beta^2 + \mu^2)}
    \right\},
  \end{split}
  \label{eq:energie}
\end{equation}
Points $z_1,\ldots,z_6$ are regular poles of second order and the order of infinity is 1 provided that $E (1 + \mu)\neq 0$. Here, the order of a singular point is defined as in \cite{Kovacic:86::}. 

The differences of exponents at the singularities in $\C$ are as follows.
\begin{equation}
  \Delta_1=\dfrac{1}{2}\sqrt{1-\dfrac{8}{\beta^2}},
  \quad \Delta_2=\dfrac{1}{2}\sqrt{1-\dfrac{8\mu^2}{\beta^2}},\quad
  \Delta_3=\Delta_4=2,\quad \Delta_5=\Delta_6=\dfrac{1}{2}.
\end{equation} 

It is evident that the differences of the exponents at singularities $z_3$ and $z_4$ are integer, and thus, in local solutions around these points, logarithmic terms may appear. Application of the method described in Appendix B verifies the presence of such terms. For singularity $z_3$, solutions of the indicial equation are $\alpha_1=3/2$ and $\alpha_2=-1/2$. The relevant one is $\alpha=\alpha_1=3/2$.  The expansion of $r(z-z_3)^2$, according to \eqref{eq:rozwr}, gives coefficients $r_0,r_1$, and $r_2$. Then $f_1,f_2$ and $g_1,g_2$ are determined from \eqref{wzfi} and \eqref{wzgi} respectively. Since $s=\alpha_1-\alpha_2=2$, the coefficient $g_2$, which multiplies the logarithm, must be found, and it is given as
\begin{equation}
  g_2=-\dfrac{\rmi (1 + \mu) ((\rmi + \beta)^2 + \mu - 2 \rmi \beta \mu -
    \mu^2)}{8 \beta (\rmi + \beta) (\beta - \rmi \mu) (1 + 
    e \beta^2 + (e - \mu) \mu - 
    \rmi \beta (1 + e (-1 + \mu) + \mu))}.
\end{equation} 
Examination of the real and imaginary part of $g_2$ yields the following conditions.
\[
\begin{split}
  &\beta (1 + \mu)[1 + \beta^4 (1 + \mu) +
  2 \beta^2 (-1 + \mu) (-1 - 2 e \mu + \mu^2) \\
  &+ \mu (-1 - 2 e (-1 + \mu) (1 + \mu^2) + \mu (-1 + \mu (-1 + (-1 +
  \mu) \mu)))]=0,\\
  &(1 + \mu)[\mu (1 + 3 \beta^2 + (-1 + \mu) \mu) (-1 + \mu^2) +
  e (\beta^6 + \mu^2 (-1 + \mu - \mu^2)\\
  & + \beta^4 (2 + \mu (-3 + 2 \mu)) + \beta^2 (1 + \mu (-1 + \mu (-1
  + (-1 + \mu) \mu))))]=0.
\end{split}
\]
This system has two solutions satisfying $\mu>0$, $\beta^2>0$ and $e\in\R$.
\begin{equation}
  \begin{split}
    \beta^2=-1 + \mu - \mu^2,\,\, e=\dfrac{1 + \mu}{1 -
      \mu}\quad \text{and}\quad \beta^2=\mu=1.
    \label{eq:waruny}
  \end{split}
\end{equation}

The condition precluding a logarithm in a local solution around $z_4$ is $g_2^\ast=0$, where ${}^\ast$
denotes complex conjugation. The corresponding solutions are the same as those given previously.

In the first solution in \eqref{eq:waruny}, condition
\begin{equation}
  \beta^2=-1 + \mu - \mu^2,
  \label{eq:warek}
\end{equation}
gives
\[
\dfrac{m_0}{m_1}=-\dfrac{\mu(\mu+1)}{\mu^2+1}\leq 0,
\]
with the only non-negative solution $m_1=0$. The second solution in \eqref{eq:waruny} also yields $m_1=0$ only.

If conditions \eqref{eq:waruny} are not satisfied, the two linearly-independent local solutions $w_1$ and $w_2$ of \eqref{eq:normal} in a neighbourhood of $z_{\ast}=z_{3}$ or $z_{\ast}=z_{4}$ have the following forms.
\begin{equation}
  \label{eq:w12}
  w_1(z) = (z - z_{\ast})^{\alpha} f(z), \qquad w_2(z)=w_1(z)\ln
  (z-z_{\ast})+(z-z_{\ast})^{\alpha-2} h(z),
\end{equation}
where $f(z)$ and $ h(z)$ are holomorphic at $z_{\ast}$ and $f(z_{\ast})\neq 0$. The local monodromy matrix, which corresponds to the continuation of the matrix of fundamental solutions along a small
loop encircling $z_{\ast}$ counter-clockwise, gives rise to a triangular monodromy matrix,
\[
\begin{bmatrix}
  -1&-2\pi\rmi\\
  0&-1
\end{bmatrix}.
\]
(For details, see \cite{mp:02::d}.) A subgroup of $\mathrm{SL}(2,\C)$ generated by a triangular non-diagonalizable matrix is not finite, and thus the differential Galois group is not finite either. Moreover, the differential Galois group $\cG$ of this equation is not any subgroup of the dihedral group because such subgroups contain only diagonalizable matrices. Thus, $\cG$ is either the full triangular group or $\mathrm{SL}(2, \C)$. But since the order of infinity is one, the necessary condition that $\cG$ is the full triangular group is not satisfied (see Lemma~\ref{lem:neces}). This means the that the differential Galois group of \eqref{eq:normal} is $\mathrm{SL}(2,\C)$, with a non-Abelian identity component equal to the whole group $\mathrm{SL}(2,\C)$.

Regarding exceptional values of energies given in \eqref{eq:energie}: the condition that the first one is real gives $\beta (1 + \mu^3 + \beta^2 (1 + \mu))=0$, which further suggests the equality \eqref{eq:warek}. Its only physical solution is $m_1$. The second expression in \eqref{eq:energie} is never real and the third one is real when \eqref{eq:warek} holds only.
\section*{Acknowledgements}
This work was partially supported by the EC Marie Curie Network for
Initial Training Astronet-II, Grant Agreement No. 289240, and 
  by the National Science Centre,  Poland,  under grant    DEC-2012/05/B/ST1/02165. 

\appendix
\section{Appendix}
Explicit form of coefficients $p_i$ in $r$ given by \eqref{eq:rr} are as follows.
\begin{equation}
 \begin{split}
&p_0=-8 e (1 + \mu),\qquad p_1=-8 (1 + \mu) (1 + 4 e (-1 + \mu) + \mu)],\\
&p_2= -8 (1 + \mu) (-4 + 4 \mu^2 + 
   e (6 + \beta^2 + 2 \mu (-5 + 3 \mu))),\\
&p_3=-48 e^2 \beta^4 - 
 16 e (-1 + \mu^2) (2 + 
    2 \beta^2 + \mu (-3 + 
       2 \mu)) - (1 + \mu)^2 (11 \beta^2\\
& + 
    24 (2 + \mu (-3 + 2 \mu))),\\
&p_4=-4 (48 e^2 \beta^4 (-1 + \mu) + (-1 + \mu) (1 + \mu)^2 (8 + 
      11 \beta^2 - 6 \mu + 8 \mu^2)\\
& + 
   2 e (1 + \mu) (1 + 
      12 \beta^4 + \beta^2 (6 + \mu (-7 + 
            6 \mu)) + \mu (-5 + \mu (5 + (-5 + \mu) \mu)))),\\
&p_5=-2 (-4 (1 + (e - \mu) \mu) (-1 + \mu^4) + 
   3 \beta^4 (9 (1 + \mu)^2 + 64 e (-1 + \mu^2)\\
& + 
      16 e^2 (3 + \mu (-8 + 
            3 \mu))) + \beta^2 (16 e (-1 + \mu^4) + 
      3 (1 + \mu)^2 (11 + \mu (-13 + 11 \mu)))),\\
&p_6=-4 \beta^2 (48 e^2 \beta^2 (\mu-1) (1 + ( \mu-5) \mu) 
+ (\mu-1) (1 + \mu)^2 (11 + 
      54 \beta^2\\
& + \mu (3 + 11 \mu)) + 
   2 e (1 + \mu) (1 + 
      3 \beta^2 (24 + \mu (-59 + 
            24 \mu))\\
& + \mu (-1 + \mu (-5 + (\mu-1) 
\mu)))),\\
&p_7=-\beta^2 (11 + 3 \beta^4 (1 + \mu)^2 + 
   12 \beta^2 (4 e (-8 + 33 \mu - 33 \mu^3 + 
         8 \mu^4) + (1 + \mu)^2 (27\\
& + \mu (-59 + 27 \mu)) + 
      4 e^2 (1 + (-2 + \mu) \mu (8 + (-14 + \mu) \mu))) +
\mu (32 + \mu (-42 \mu\\
& + \mu^3 (32 + 11 \mu) - 
         32 e (-1 + \mu^2)))),\\
&p_8=-4 \beta^2 (3 \beta^4 (-1 + \mu) (1 + \mu)^2 - 
   2 (1 + (e - \mu) \mu) (\mu + \mu^4)\\
& + 
   2 \beta^2 (-24 e^2 (-1 + \mu) \mu (1 + (-5 + \mu) \mu) + 
      27 (-1 + \mu) (1 + \mu)^2 (1 + (-3 + \mu) \mu)\\
& + 
      2 e (1 + \mu) (6 + \mu (-76 + \mu (157 - 76 \mu + 
               6 \mu^2))))),\\
&p_9=-2 \beta^4 (27 + 
   3 \beta^2 (1 + \mu)^2 (3 + \mu (-5 + 
         3 \mu)) + \mu (-176 - 
      140 e (-1 + 4 \mu - 4 \mu^3\\
& + \mu^4) + 
      48 e^2 \mu (3 + \mu (-8 + 
            3 \mu)) + \mu (4 + \mu (414 + \mu (4 + \mu 
(-176 + 27 \mu)))))),\\
&p_{10}=-4 \beta^4 (3 \beta^2 (-1 + \mu^2 - \mu^3 + \mu^5) - 
   2 \mu (1 + (e - \mu) \mu) (-11 + \mu (24 \\
&+ 
         24 e (\mu-1) + (24 - 11 \mu) \mu))),\\
&p_{11}=-3 \beta^4 (16 \mu^2 (1 + (e - \mu) \mu)^2 + \beta^2 (1 +
\mu^3)^2)
\end{split}.
\label{eq:coeffy}
\end{equation} 
\section{Second-order differential equation in reduced form with rational coefficients and its differential Galois group}
Consider the second-order linear differential equation in reduced form with rational coefficient
\begin{equation}
  w'' =r(z)w,\qquad r(z)\in\mathbb{C}(z).
  \label{pom}
\end{equation}
A point $z=c\in\mathbb{C}$ is a singular point of this equation if it is a pole of $r(z)$. A singular point is a regular point if at this point $(z-c)^2r(z)$ is holomorphic.

Assume that $c$ is a regular point and look for local solutions of this equation in a neighbourhood of this point. For simplicity, assume that $c=0$. Look for a solution of the following form (see
e.g.~\cite{Whittaker:35::}),
\begin{equation}
  w=z^{\alpha}f(z),\qquad f(z)=1+\sum_{i=1}^{\infty}f_i z^i,
  \label{local}
\end{equation}
where $\alpha,f_i$ for $i\in\mathbb{N}$ are constants to be determined. Substituting in \eqref{pom} yields
\begin{equation}
  z^{\alpha-2}\left[\alpha(\alpha-1)+\sum_{i=1}^{\infty}
    (\alpha+i)(\alpha+i-1)f_iz^i\right]
  -r(z)z^{\alpha}
  \left(1+\sum_{i=1}^{\infty}f_iz^i\right)=0.
\end{equation}
Multiplying by $z^2$, apply Taylor's formula for the analytic function
$z^2r(z)$
\begin{equation}
  z^2r(z)=r_0+\sum_{i=1}^{\infty} r_iz^i,
  \label{eq:rozwr}
\end{equation} 
and equating to zero the coefficients of successive powers of $t$, obtain the sequence of equations determining unknown coefficients in series \eqref{local}. The first of these equations, called the
indicial equation and determining $\alpha$, has the following form.
\begin{equation}
  \alpha(\alpha-1)-r_0=0.
\end{equation}
From the remaining equations $f_i$ is evaluated.
\begin{equation}
  f_1=\dfrac{r_1}{2\alpha},\qquad
  f_i=\dfrac{1}{i(i+2\alpha-1)}\left(\sum_{j=1}^{i-1}r_jf_{i-j}+r_i\right),\quad
  \text{for}\quad
  i\geq 2.
  \label{wzfi}
\end{equation}
Let $\alpha_1$ and $\alpha_2$ be the roots of the indical equation. If the difference $s=\alpha_1-\alpha_2$ is not  an integer, then equation~\eqref{pom}  has two local solutions of the form
\begin{equation}
  w_1=z^{\alpha_1}\left(1+\sum_{i=1}^{\infty}f_i z^i\right), \qquad
  w_2=z^{\alpha_2}\left(1+\sum_{i=1}^{\infty}f_i' z^i\right).
  \label{eq:2rozw}
\end{equation}
In the case where $s$ is a positive integer or zero, the series $w_2$ may not exist or coincide with $w_1$, and a second independent solution must be constructed \cite{Whittaker:35::} as
\begin{equation}
  w_2(z)=w_1(z)\int^z\dfrac{\mbox{d}x}{w_1(x)^2}=
  w_1(z)\int^zz^{-2\alpha_1}\dfrac{\mbox{d}x}{f(x)^2}=
  w_1(z)\int^zz^{-s-1}\dfrac{\mbox{d}x}{f(x)^2}.
\end{equation}
In the last equality the relation $2\alpha_1=s+1$ is applied.

Denoting
\begin{equation}
  \dfrac{1}{f(z)^2}=1+\sum_{i=1}^{\infty}g_iz^i,
\end{equation}
the second solution is written as
\begin{equation}
  w_2(z)=w_1(z)g_s\ln
  z+z^{\alpha_2}\left(-\dfrac{1}{s}+\sum_{i=1}^{\infty}h_it^i\right),
  \label{pp}
\end{equation}
where $h_i$ are constants.

Coefficients $g_i$ are given by
\begin{align}
  g_1&=-2f_1,\qquad
  g_2=-2f_2-2f_1g_1-f_1^2=-2f_2+3f_1^2,\shortintertext{and}
  g_i&=-2f_i-\sum_{j=1}^{i-1}(2g_{i-j}+f_{i-j})f_j-\sum_{j=1}^{i-2}f_j\sum_{k=1}^{
    i-j-1} f_kg_{i-j-k},
  \label{wzgi}
\end{align}
for $ i\geq 3$, while coefficients $h_i$ are defined as follows.
\begin{equation}
  h_i=\begin{dcases}
    -\dfrac{f_1}{s}-\dfrac{g_1}{s-1}& \mtext{for}
    i=1,\\
    -\dfrac{f_i}{s}-\dfrac{g_i}{s-i}-\sum_{j=1}^{i-1}\dfrac{g_j}{s-j}f_{i-j}
    & \mtext{for}
    1<i<s,\\
    -\dfrac{f_s}{s}& \mtext{for}  i=s,\\
    -\dfrac{f_{s+1}}{s}+g_{s+1}-\sum_{j=1}^{s-1}\dfrac{g_j}{s-j}f_{s+1-j}
    & \mtext{for}
    i=s+1,\\
    -\dfrac{f_i}{s}+\dfrac{g_i}{i-s}-\sum_{j=1}^{s-1}\dfrac{g_j}{s-j}f_{i-j}+
    \sum_{j=s+1}^{i-1}\dfrac{g_j}{j-s}f_{i-j}& \mtext{for}
    i\geq s+2.
  \end{dcases}
\end{equation}

The differential Galois group of equation~\eqref{pom} is a subgroup of  $\mathrm{SL}(2,\C)$. Classification of these subgroups is well known. 
The following lemma shows that the differential Galois group determines the form of solutions  (see e.g.~\cite{Kovacic:86::,Morales:99::c}).
\begin{lemma}
  \label{lem:alg}
  Let $\cG$ be the differential Galois group of equation~\eqref{eq:normal}. Then one of four cases may occur.
  \begin{description}
  \item[Case I] $\cG$ is conjugate to a subgroup of the triangular
    group,
    \[
    \cT = \left\{ \begin{bmatrix} a & b\\
        0 & a^{-1}
      \end{bmatrix} \; \biggl| \; a\in\C^*, b\in\C\right\} text{;}
    \]
    in this case equation \eqref{eq:normal} has an exponential solution of the form $y=P\exp\int\omega$, where $P\in\C[z]$ and $\omega\in\C(z)$.
  \item[Case II] $\cG$ is conjugate to a subgroup of
    \[
    \cD^\dag = \left\{ \begin{bmatrix} c & 0\\
        0 & c^{-1}
      \end{bmatrix} \; \biggl| \; c\in\C^*\right\} \cup
    \left\{ \begin{bmatrix} 0 & c\\
        -c^{-1} & 0
      \end{bmatrix} \; \biggl| \; c\in\C^*\right\}text{;}
    \]
    in this case equation \eqref{eq:normal} has a solution of the form $y=\exp\int \omega$, where $\omega$ is algebraic over $\C(z)$ of degree 2,
  \item[Case III] $\cG$ is primitive and finite; in this case all solutions of equation~\eqref{eq:normal} are algebraic, thus $y=\exp\int\omega$, where $\omega$ belongs to an algebraic extension of ${\C}(z)$ of degree $n=4,6$ or 12.

  \item[Case IV] $\cG= \mathrm{SL}(2,\C)$ and equation \eqref{eq:normal} has no Liouvillian solution.
  \end{description}
\end{lemma}

Kovacic \cite{Kovacic:86::} formulates the necessary conditions for the respective cases from Lemma~\ref{lem:alg} to hold.

Write $r(z)\in\mathbb{C}(z)$ in the form
\begin{equation*}
  r(z) = \frac{s(z)}{t(z)}, \qquad s(z),\, t(z) \in \mathbb{C}[z],
\end{equation*}
where $s(z)$ and $t(z)$ are relatively prime polynomials and $t(z)$ is monic. The roots of $t(z)$ are poles of $r(z)$. Denote $\Sigma':= \{c\in\mathbb{C}\,\vert\, t(c) =0 \}$ and $\Sigma:=\Sigma'\cup\{\infty\}$. The order $\mathrm{ord}(c)$ of $c\in\Sigma'$ is equal to the multiplicity of $c$ as a root of $t(z)$, the order of infinity is defined by
\[
\mathrm{ord}(\infty):=\deg t - \deg s.
\]

\begin{lemma}
  \label{lem:neces}
  The necessary conditions for the respective cases in Lemma~\ref{lem:alg} are the following.
  \begin{description}
  \item[Case~I.] Every pole of $r$ must have even order or else have order~1. The order of $r$ at $\infty$ must be even or else be greater than~2.
  \item[Case~II.] $r$ must have at least one pole that either has odd order greater than~2 or else has order~2.
  \item[Case~III.] The order of a pole of $r$ cannot exceed~2 and the order of $r$ at $\infty$ must be at least 2. If the partial-fraction expansion of $r$ is
    \begin{equation}
      \label{eq:r}
      r(z)=\sum_{i}\dfrac{a_i}{(z-c_i)^2}+
      \sum_{j}\dfrac{b_j}{z-d_j},
    \end{equation}
    then $\Delta_i=\sqrt{1+4a_i}\in\mathbb{Q}$, for each $i$, $\sum_{j}b_j=0$ and if
    \begin{equation*}
      g=\sum_ia_i+\sum_jb_jd_j,
    \end{equation*}
    then $\sqrt{1+4g}\in\mathbb{Q}$.
  \end{description}
  \label{kovacic}
\end{lemma}

\newcommand{\noopsort}[1]{}\def\polhk#1{\setbox0=\hbox{#1}{\ooalign{\hidewidth
  \lower1.5ex\hbox{`}\hidewidth\crcr\unhbox0}}} \def\cprime{$'$}
  \def\cydot{\leavevmode\raise.4ex\hbox{.}} \def\cprime{$'$}
  \def\polhk#1{\setbox0=\hbox{#1}{\ooalign{\hidewidth
  \lower1.5ex\hbox{`}\hidewidth\crcr\unhbox0}}} \def\cprime{$'$}
  \def\cprime{$'$} \def\cprime{$'$} \def\cprime{$'$}

\end{document}